\documentclass[12pt]{article}
\usepackage{epsfig}

\def\beq{\begin{equation}}
\def\eeq#1{\label{#1}\end{equation}}
\def\eeqn{\end{equation}}

\def\beqa{\begin{eqnarray}}
\def\eeqa#1{\label{#1}\end{eqnarray}}
\def\eeqan{\end{eqnarray}}

\let\bar=\overbar

\def\Dslash{\not{\hbox{\kern-4pt $D$}}}
\def\dslash{\not{\hbox{\kern-2pt $\del$}}}

\def\msb{{\bar{\ssstyle M \kern -1pt S}}}

\def\Title#1{\begin{center} {\Large {\bf #1} } \end{center}}

\begin{document}

\Title{Lifetimes and oscillations of heavy mesons}

\bigskip\bigskip

%+\addtocontents{toc}{{\it D. Reggiano}}
%+\label{ReggianoStart}

\begin{raggedright}  

{\it Alexander Lenz \index{Lenz, A.}\\
Institut f{\"u}r theoretische Physik\\
Universit{\"a}t Regensburg\\
D-93040 Regensburg - Germany}
\bigskip\bigskip
\end{raggedright}

\section{Abstract}

We review the theoretical status of the lifetime ratios 
$\tau_{B^+} / \tau_{B_d}$ and $\tau_{B_s} / \tau_{B_d}$ and of the mixing 
quantities $\Delta M_s$, $\Delta \Gamma_s$ and $\phi_s$.
We show that the ratio $\Delta \Gamma_s / \Delta M_s$ can be determined 
with almost no non-perturbative uncertainties.
Finally we explain how this precise determination of the standard model
values can be used to find possible new physics contributions in 
$\Delta M_s $, $\Delta \Gamma_s$, $\Delta \Gamma_s / \Delta M_s$ 
and $a_{fs}^s$. Combining the latest experimental bounds on these 
quantities one already gets  some hints for new physics contributions.

\section{Introduction}
Inclusive decays 
(see e.g. \cite{inclusive} or \cite{CERN} and references therein) 
and lifetimes of heavy mesons can be calculated within the framework of the
so-called heavy quark expansion (HQE) \cite{HQE1,HQE2}. In this approach
the decay rate is calculated in an expansion in inverse powers of the heavy 
b-quark mass.
\begin{equation}
\Gamma = 
\Gamma_0 + \frac{\Lambda^2}{m_b^2} \Gamma_2
         + \frac{\Lambda^3}{m_b^3} \Gamma_3 + \ldots
\end{equation}
$\Gamma_0$ represents the decay of a free heavy b-quark, 
according to this contribution all b-mesons have the same lifetime.
The first correction arises at order $1/m_b^2$, they are due to the 
kinetic and the chromomagnetic operator. At order $1/m_b^3$ the 
spectator quark gets involved in the weak 
annihilation and Pauli interference diagrams \cite{HQE1,spectator}. 
This contributions are 
numerically enhanced by a phase space factor of $16 \pi^2$.
Each of the $\Gamma_i$ contains perturbatively calculable Wilson 
coefficients and non-perturbative parameters, like decay constants or
bag parameters. 
Unfortunately the theoretical predictions for the decay
constants vary over a wide range: quenched lattice determinations for 
$f_{B_s}$  tend to give values of ${\cal O} (200)$ MeV, while recent
unquenched calculations with 2+1 dynamical light flavors give values around
260 MeV - for a more detailed discussion see \cite{Vittorio, LN, Onogi}. 
Since lifetime differences depend quadratically on the decay 
constants, going from 200 MeV to 260 MeV results in an increase of $70\%$.
Here clearly theoretical progress is necessary to pin down the error on 
the decay constants considerably.

In view of several new theoretical and experimental developments
we update the numbers present in the literature (see e.g. \cite{updates}).

\section{Lifetimes}
The lifetime ratio of two heavy mesons can be written as
\begin{equation}
\frac{\tau_1}{\tau_2} = 1 +
\frac{\Lambda^3}{m_b^3} 
\left( \Gamma_3^{(0)} + \frac{\alpha_s}{4 \pi} \Gamma_3^{(1)} + \ldots\right) +
\frac{\Lambda^4}{m_b^4} 
\left( \Gamma_4^{(0)} + \frac{\alpha_s}{4 \pi} \Gamma_4^{(1)} + \ldots\right) +
\ldots
\end{equation}
If one neglects small isospin or SU(3) violating effects one 
has no $1/m_b^2$ corrections \footnote{In the case of 
$\tau_{\Lambda_b}/ \tau_{B_d}$ these effects are expected to be of the order
of $5\%$.} and a deviation of the lifetime ratio from
one starts at order $1/m_b^3$.

\subsection{$\tau_{B^+}/ \tau_{B_d}$}

The leading term $\Gamma_3^{(0)}$ has been determined in \cite{HQE1, tauLO}.
For a quantitative treatment of the lifetime ratios NLO QCD corrections 
are mandatory -  $\Gamma_3^{(1)}$ has been determined in 
\cite{BBGLN02, rome02}. Subleading effects of ${\cal O} (1/m_b)$ turned 
out to be negligible \cite{lifetime1m}.
The matrix elements of the arising four-quark operators have been 
determined in \cite{lifetimelattice}.
Using the result from \cite{BBGLN02}
 \begin{eqnarray} 
 \label{eqBBGLN02}
 \lefteqn{
  \frac{\tau(B^+)}{\tau(B_d^0)} - 1
  \; = \;
  \tau(B^+) \, \left[ \Gamma (B_d^0) - \Gamma (B^+) \right]}
      \\ & = &
      0.0325 \, \frac{\tau(B^+)}{1.653 \, \mbox{ps}}
        \left( \frac{|V_{cb}|}{0.04} \right)^2
        \, \left( \frac{m_b}{4.8\mbox{GeV}} \right)^2 \,
           \left( \frac{f_B}{200\mbox{MeV}} \right)^2 \,
        × \nonumber \\ &&
      \Big[ \, ( 1.0 \pm 0.2) \, B_1 \; + \; (0.1\pm 0.1) \, B_2 \;
      - \;
         (18.4\pm 0.9) \, \epsilon_1 \; + \; (4.0\pm 0.2) \, \epsilon_2
      \, \Big] + \delta_{1/m} \nonumber
\end{eqnarray}
one gets with the matrix elements from Becirevic \cite{lifetimelattice}
$( B_1 = 1.10 \pm 0.20 ;\,  B_2 = 0.79 \pm 0.10; \, 
   \epsilon_1 = -0.02 \pm 0.02  ; \, 
   \epsilon_2 =  0.03 \pm 0.01)$ 
and the values $V_{cb} = 0.0415$, $m_b = 4.63$ GeV and $f_{B} = 216 $ MeV
\cite{fB}:
\begin{equation}
\left[ \frac{\tau(B^+)}{\tau(B_d^0)} \right]_{\rm NLO} \; = \;
1.063\pm 0.027 \, ,
\end{equation}
which is in excellent agreement with the experimental number \cite{PDG, ay}
\begin{equation}
\left[ \frac{\tau(B^+)}{\tau(B_d^0)} \right] \; = \;
1.071\pm 0.009 \, .
\end{equation}
From Eq. (\ref{eqBBGLN02}) one clearly sees that a precise knowledge of
the color octet bag parameters $\epsilon_1$ and $\epsilon_2$ - 
these parameters are of order $1/N_c$ - is mandatory 
since their coefficients are numerically enhanced. 
Here clearly more work has to be done.
\subsection{$\tau_{B_s}/ \tau_{B_d}$}

In the lifetime ratio $\tau_{B_s}/ \tau_{B_d}$ a cancellation of 
weak annihilation contributions arises, that differ only by small
SU(3)-violation effects. One expects a number that is close to one
\cite{tauLO, rome02, tauBs, BBD}
\begin{equation}
\frac{\tau(B_s)}{\tau(B_d)}  \; = \; 1.00 \pm 0.01 \, .
\end{equation}
This expectation is confirmed by experiment \cite{HFAG,ay}
\begin{equation}
\frac{\tau(B_s)}{\tau(B_d)}  \; = \; 0.957 \pm 0.027\, ,
\end{equation}
although more precise experimental numbers would be very desirable.
\subsection{$\tau_{B^+_c}$}
The lifetime of the doubly heavy meson $B_c$ has been investigated in
\cite{taubctheory}
\begin{equation}
\tau(B_c)  \; = \; 0.52_{-0.12}^{+0.18} \,  \mbox{ps} \, .
\end{equation}
In addition to the b-quark now also the c-charm quark can decay, giving rise
to the biggest contribution to the total decay rate.
The current experimental number  \cite{taubcexp}
\begin{equation}
\tau(B_c)  \; = \; 0.469 \pm 0.027 \, \mbox{ps}
\end{equation}
agrees nicely with the theoretical prediction, but it has much smaller 
errors.
Here clearly some theoretical improvements are necessary to pin down the error.

\section{Mixing Parameters}

In this section we briefly investigate the status of the mixing parameters.
For a more detailed review we refer the interested reader to \cite{LN}.
\\
The mixing of the neutral B-mesons is described by the off diagonal
elements $\Gamma_{12}$ and $M_{12}$ of the mixing matrix.
$\Gamma_{12}$ stems from the absorptive part of the box diagrams - only
internal up and charm quarks contribute, while $M_{12}$ stems from
the dispersive part of the box diagram, therefore being sensitive to 
heavy internal 
particles like the top quark or heavy new physics particles.
By diagonalizing the mixing matrix we obtain the physical eigenstates
$B_H$ and $B_L$ with defined masses ($M_H, M_L$) and defined decay rates
($\Gamma_H, \Gamma_L$) in terms of the flavor eigenstates 
$B_s = (\bar b s)$ and  $\bar B_s =  (b \bar s)$:
\begin{equation}
B_H  := p \; B + q \; \bar{B} \, ,
\hspace{1cm} 
B_L  :=  p \; B - q \; \bar{B}
\hspace{1cm} 
\mbox{with} 
\, \, \,  
|p|^{2} + |q|^{2}  =  1\, .
\end{equation}
The calculable quantities {$  |M_{12}|$}, {$ |\Gamma_{12}|$} and 
{$ \phi = \mbox{arg}( -M_{12}/\Gamma_{12})$}
can be related to three observables:
\begin{itemize}
\item \underline{\bf Mass difference:} 
\begin{equation}
 \Delta M := M_H - M_L  = 
        2 { |M_{12}|} \left( 1 +
        { \frac{1}{8} 
        \frac{|\Gamma_{12}|^2}{|M_{12}|^2} \sin^2 \phi + ...}\right) 
\end{equation}
        ${ |M_{12}|}$ is due to heavy internal particles in the boxdiagrams
        like the top-quark or SUSY-particles.
\item \underline{\bf Decay rate difference:} 
      \begin{equation}
       \Delta \Gamma := \Gamma_L - \Gamma_H = 
        2 { |\Gamma_{12}| \cos  \phi }
        \left( 1 -
        { \frac{1}{8} \frac{|\Gamma_{12}|^2}{|M_{12}|^2} \sin^2 \phi 
        + ...}\right)
      \end{equation}
      ${ |\Gamma_{12}|}$ is due to light internal particles: particles,
      like the up- and the charm-quark. It is therefore very insensitive 
      to new physics contributions.
\item \underline{\bf {Flavor specific}/semileptonic CP asymmetries:}
      \\
      A decay $B_q \to f$ is called flavor specific, if the 
      decays $\bar{B}_q \to f$ and { $B_q \to \bar{f}$} are forbidden
      and if no direct CP violation occurs, i.e. 
      { $|\langle f | B_q \rangle| =| \langle\bar{f} | \bar{B}_q\rangle|$}.
      Some examples are { $B_s \to D_s^- \pi^+$} or { $B_q \to X l \nu$}
      (therefore the name semileptonic CP asymmetry).
      The flavor specific CP asymmetry is defined as
\begin{eqnarray}
a_{fs} & =  &
\frac{\Gamma \left( \bar{B}_q (t) \to      f  \right) 
    - \Gamma \left(     {B}_q (t) \to \bar{f} \right)}
     {\Gamma \left( \bar{B}_q (t) \to      f  \right) 
    + \Gamma \left(     {B}_q (t) \to \bar{f} \right)}
= 
 - 2 \left( \left| \frac{q}{p}\right| - 1 \right) 
\nonumber \\
& = & { \mbox{Im}
 \frac{\Gamma_{12}}{M_{12}}} =
\frac{\Delta \Gamma}{\Delta M} \tan \phi \, .
\nonumber
\end{eqnarray}
\end{itemize}

\subsection{Mass difference}

Calculating the box diagram with internal top quarks one obtains
 \begin{eqnarray}
        M_{12,q} & = & \frac{G_F^2}{12 \pi^2} 
          (V_{tq}^* V_{tb})^2 M_W^2 S_0(x_t)
          {B_{B_q} f_{B_q}^2  M_{B_q}} \hat{\eta }_B
        \end{eqnarray}
The Inami-Lim function $S_0 (x_t = \bar{m}_t^2/M_W^2)$ 
\cite{IL} is the result of the box diagram 
without any gluon corrections. The NLO QCD correction is parameterized by 
$\hat{\eta}_B \approx 0.84$ \cite{BJW}.
The non-perturbative matrix element of the operator 
$Q= (\bar{b}q)_{V-A} (\bar{b}q)_{V-A}$ is parameterized by the 
bag parameter $B$ and the decay constant $f_B$
\begin{equation}
      \langle \bar{B_q}|Q |B_q \rangle
=
 \frac{8}{3} {B_{B_q} f_{B_q}^2  M_{B_q}}\, .
\end{equation}
Using the conservative estimate $f_{B_s} = 240 \pm 40$ MeV \cite{LN} 
and the bag parameter $B$ from JLQCD \cite{JLQCD} we obtain
\begin{equation}
\Delta M_s = 19.3 \pm 6.4 \pm 1.9 \, \mbox{ps}^{-1}
\end{equation}
The first error stems from the uncertainty in $f_{B_s}$ and the second error
summarizes the remaining theoretical uncertainties.
The determination of $\Delta M_d$ is affected by even larger 
uncertainties because here one has to extrapolate to the small mass
of the down-quark.
The ratio $\Delta M_s / \Delta M_d$ is theoretically better under
control since in the ratio of the non-perturbative parameters many systematic
errors cancel.

This year also $\Delta M_s$ was measured, leading to the pleasant situation
of having very precise experimental numbers at hand
\cite{HFAG, menzemer, deltamsexp} 
\begin{eqnarray}
\Delta M_d & = & 0.507 \pm 0.004 \, \mbox{ps}^{-1} \, ,
\\
\Delta M_s & = & 17.77 \pm 0.12  \, \mbox{ps}^{-1} \, .
\end{eqnarray}
To be able to distinguish possible new physics contributions to $\Delta M_s$
from QCD uncertainties much more precise numbers for $f_{B_s}$ are needed.

\subsection{Decay rate difference and flavor specific CP asymmetries}

In order to determine the decay rate difference of the neutral B-mesons 
and flavor specific CP asymmetries a precise determination of $\Gamma_{12}$ 
is needed.
With the help of the HQE $\Gamma_{12}$ can be written as
\begin{equation}
\Gamma_{12} = 
\frac{\Lambda^3}{m_b^3} \left( \Gamma_3^{(0)} + \frac{\alpha_s}{4 \pi} \Gamma_3^{(1)} + \ldots\right) +
\frac{\Lambda^4}{m_b^4} \left( \Gamma_4^{(0)} + \frac{\alpha_s}{4 \pi} \Gamma_4^{(1)} + \ldots\right) + \ldots
\end{equation}
The leading term $\Gamma_3^{(0)}$ was determined in \cite{dgLO}.
The numerical and conceptual important NLO-QCD corrections ($\Gamma_3^{(1)}$) 
were determined in \cite{BBGLN98,BBLN03, rome03}.
Subleading $1/m$-corrections, i.e. $\Gamma_4^{(0)}$ were calculated 
in \cite{BBD, 1overm}
and even the Wilson coefficients of the $1/m^2$-corrections 
($\Gamma_5^{(0)}$) were calculated and found to be small \cite{LN2}.

Besides the already known operator $Q$ in the calculation of $\Gamma_{12}$
three additional operators arise $\tilde Q, Q_S$ and $ \tilde Q_S$. The tilde 
stands for a color rearrangement and index $S$ corresponds to a S-P Dirac 
structure instead of the V-A-structure.
$Q = \tilde Q$ and it can be shown \cite{BBD, BBGLN98, LN} that a certain 
combination of $Q, Q_S$ and $\tilde Q_S$ is suppressed by powers of $1/m_b$
and $\alpha_s$
\begin{eqnarray}
\label{R0}
 \tilde Q = Q & \mbox{ and} & 
      R_0 = Q_S + \alpha_1 \tilde Q_S + \frac{\alpha_2}{2} Q 
      = {\cal O} \left( \frac{1}{m_b}, \alpha_s \right)
\end{eqnarray}
with $\alpha_i = 1 + {\cal O} \left( \alpha_s \right)$, for more details 
see \cite{LN}.
In the literature \cite{dgLO, BBD,BBGLN98,BBLN03, rome03} always 
$\tilde{Q}_S$ was eliminated - with the help of Eq. (\ref{R0}) - and
one was left with the operator basis $\{ Q, Q_S \}$, 
which we call in the following the {\it old basis}.
Working in the old basis one finds several serious drawbacks:
\begin{itemize}
\item An almost complete cancellation of the coefficient of the operator
      $Q$ takes place, while the operator $Q_S$ is dominant. So in the ratio
      $\Delta \Gamma_s / \Delta M_s$ the only coefficient that is free 
      of non-perturbative uncertainties is numerically negligible.
\item The $1/m$ corrections are abnormally large - all contributions have
      the same sign.
\item The $\alpha_s$-corrections and the remaining $\mu$-dependence is 
      unexpectedly large.
\end{itemize}
In \cite{LN} it was found, that expressing $\Gamma_{12}$ in terms 
of the {\it new basis}  $\{ Q, \tilde Q_S \}$ one gets a result, 
that is free of the above shortcomings.
The change of the basis corresponds to throwing away
certain contributions of ${\cal O} (\alpha_s^2)$ and 
${\cal O} (\alpha_s / m_b)$, which is beyond the calculated accuracy.
For our new determination of $\Gamma_{12}$ we also use 
the $\overline{\mbox{MS}}$-scheme \cite{MS}, besides the pole scheme  
for the b-quark mass. Moreover we sum up logarithms of the form
$z \ln z$ -  with $z = m_c^2/m_b^2$ - to all orders, following \cite{BBGLN02}
and of course we have to include also subleading CKM-structures to determine
$a_{fs}$, as done in \cite{BBLN03, rome03}.

In the old basis one obtains
\begin{eqnarray}
\Delta \Gamma_s & = & 
\left( \frac{f_{B_s}}{240 \, \mbox{MeV}} \right)^2
\left[ { 0.002 B} + { 0.094  B_S'} - 
\left(0.033  B_{\tilde{R}_2} +  { 0.019 B_{R_0}}
+  0.005 B_R \right) \right] \, ,
\nonumber
\\
&&
\\
\frac{\Delta \Gamma_s}{\Delta M_s}  & = & \, \, \, \,
\, \, \,  \, \, \, \, \, \, \, \,  \, \, \, \, 10^{-4} \cdot
\left[ { 0.9}  \, \, \, \,  \, \, \, \,  \, \, \, \,
+ { 40.9  \frac{ B_S'}{B}}  
- \left(14.4 \frac{B_{\tilde{R}_2}}{B} + { 8.5 \frac{B_{R_0}}{B}} 
+ 2.1 \frac{B_R}{B}  \right) \right] \, ,
\label{oldbasis}
\end{eqnarray}
with
\begin{equation}
      \langle \bar{B_s}|Q_S |B_s \rangle
=
 - \frac{5}{3} {B_{S}' f_{B_s}^2  M_{B_s}}\, , \, \,\,\,\,B_X' := B_X 
\frac{ M_{B_s}^2}{\left( \bar{m}_b + \bar{m}_s  \right)^2} \, .
\end{equation}
In Eq. (\ref{oldbasis}) we have explicitly shown the dependence 
on the dominant $1/m$ operators $R_2$ and $R_0$ 
(see \cite{BBD,LN} for the definition). The remaining power corrections
are summarized in the coefficient of $B_R$. One clearly sees that 
the cancellation in the coefficient of $B$ leads to the undesirable 
situation, that the only coefficient in $\Delta \Gamma / \Delta M$ that
is free of non-perturbative uncertainties is negligible.

\begin{figure}[htb]
\begin{center}
\epsfig{file=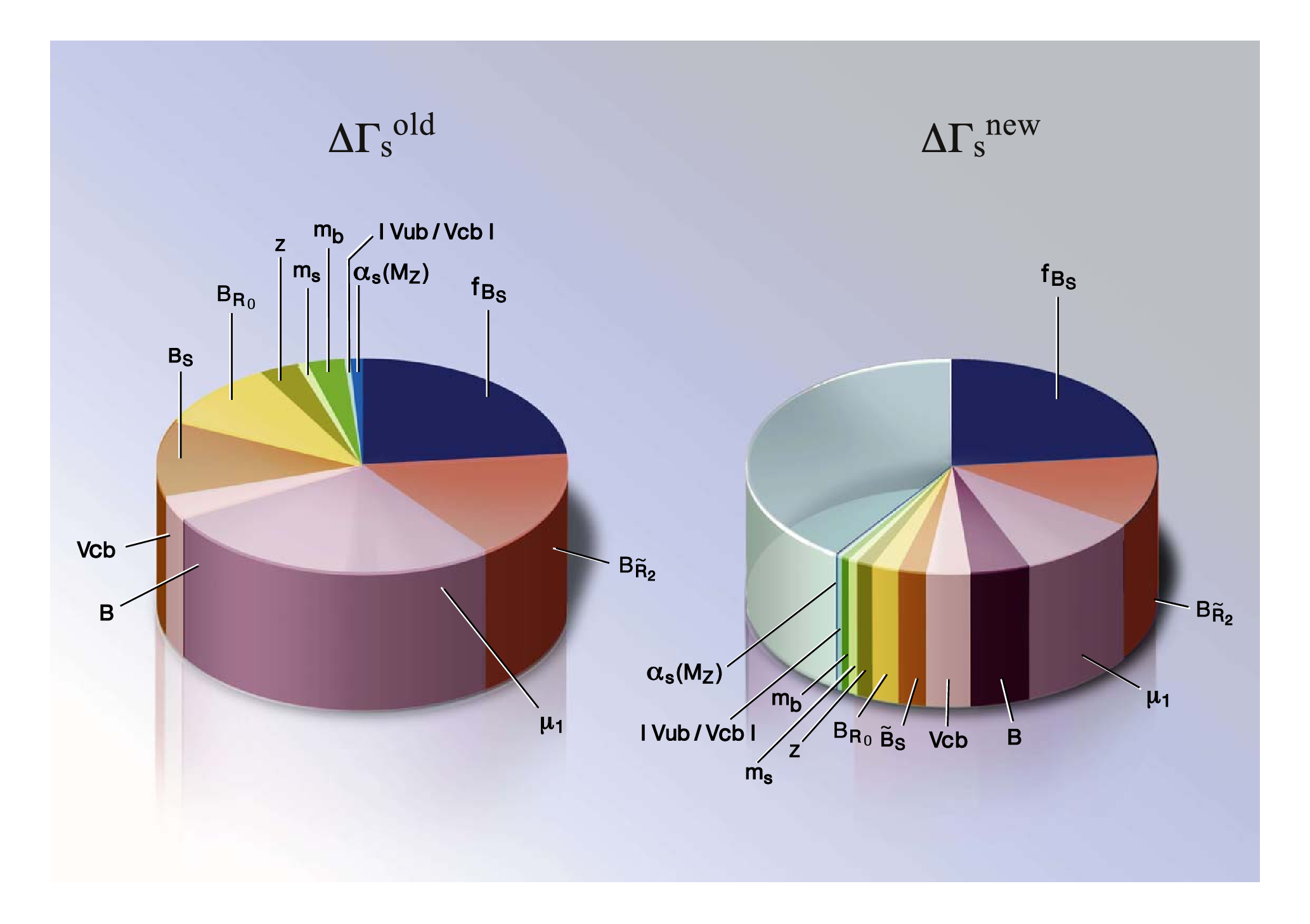,height=4 in}
\caption{Error budget for the determination of $\Delta \Gamma_s$ in the old and the new basis.}
\label{fig:Kuchen}
\end{center}
\end{figure}

This changes however dramatically if one uses the  new basis
\begin{eqnarray}
\Delta \Gamma_s & = & 
\left( \frac{f_{B_s}}{240 \, \mbox{MeV}} \right)^2
\left[ { 0.105 B} + { 0.024  \tilde B_S'} - 
\left(0.030  B_{\tilde{R}_2} -  { 0.006 B_{R_0}}
+  0.003 B_R \right) \right]
\nonumber
\\
&&
\label{dgnew}
\\
\frac{\Delta \Gamma_s}{\Delta M_s}  & = & \, \, \, \,
\, \, \,  \, \, \, \, \, \, \, \,  \, \, \, \, 10^{-4} \cdot
\left[ { 46.2}  \, \,  \, \, \, \,  \, \, \, \,
+ { 10.6 \frac{ B_S'}{B}}  
- \left(13.2 \frac{B_{\tilde{R}_2}}{B} - { 2.5 \frac{B_{R_0}}{B}} 
+ 1.2 \frac{B_R}{B}  \right) \right] 
\label{dgdmnew}
\end{eqnarray}
with
\begin{equation}
      \langle \bar{B_s}|\tilde{Q}_S |B_s \rangle
=
\frac{1}{3} {\tilde{B}_{S}' f_{B_s}^2  M_{B_s}}\, .
\end{equation}
Now the dominant part of $\Delta \Gamma / \Delta M $ can be determined without 
any hadronic uncertainties!
\\
Using the non-perturbative parameters from
\cite{JLQCD, Becirevic}, we obtain the following final numbers
(see \cite{LN} for the complete list of the numerical input parameters)
\begin{eqnarray}
\Delta \Gamma_s & = & \left( 0.096   \pm 0.039 \right) \, \mbox{ps}^{-1}
 \hspace{0.25cm} \Rightarrow \hspace{0.25cm}
\frac{\Delta \Gamma_s}{\Gamma_s}  = \Delta \Gamma_s \cdot
 \tau_{B_d} = 0.147 \pm 0.060 \label{findg}\, ,
\\
a_{fs}^s & = & \left( 2.06 \pm 0.57 \right) \cdot 10^{-5}\, ,
\\
\frac{\Delta \Gamma_s}{\Delta M_s}  & = & 
\left( 49. 7 \pm 9.4 \right) \cdot 10^{-4}\, ,
\\
\phi_s & = & 0.0041 \pm 0.0008 \, \, \, = \, \, \, 0.24^\circ \pm 0.04 \, .
\label{finphi}
\end{eqnarray}
The composition of the theoretical error of $\Delta \Gamma$ is compared 
for the use of the old and the new basis in Fig. (\ref{fig:Kuchen}).
The by far dominant error comes from the decay constant $f_{B_s}$, followed 
by the uncertainty due to the power suppressed operator $\tilde{R}_2$
and the remaining $\mu$-dependence.
In this case the theoretical improvement due to the change 
of basis is somehow limited by the huge uncertainty due to $f_{B_s}$, 
which is the same in both bases.

\begin{figure}[htb]
\begin{center}
\epsfig{file=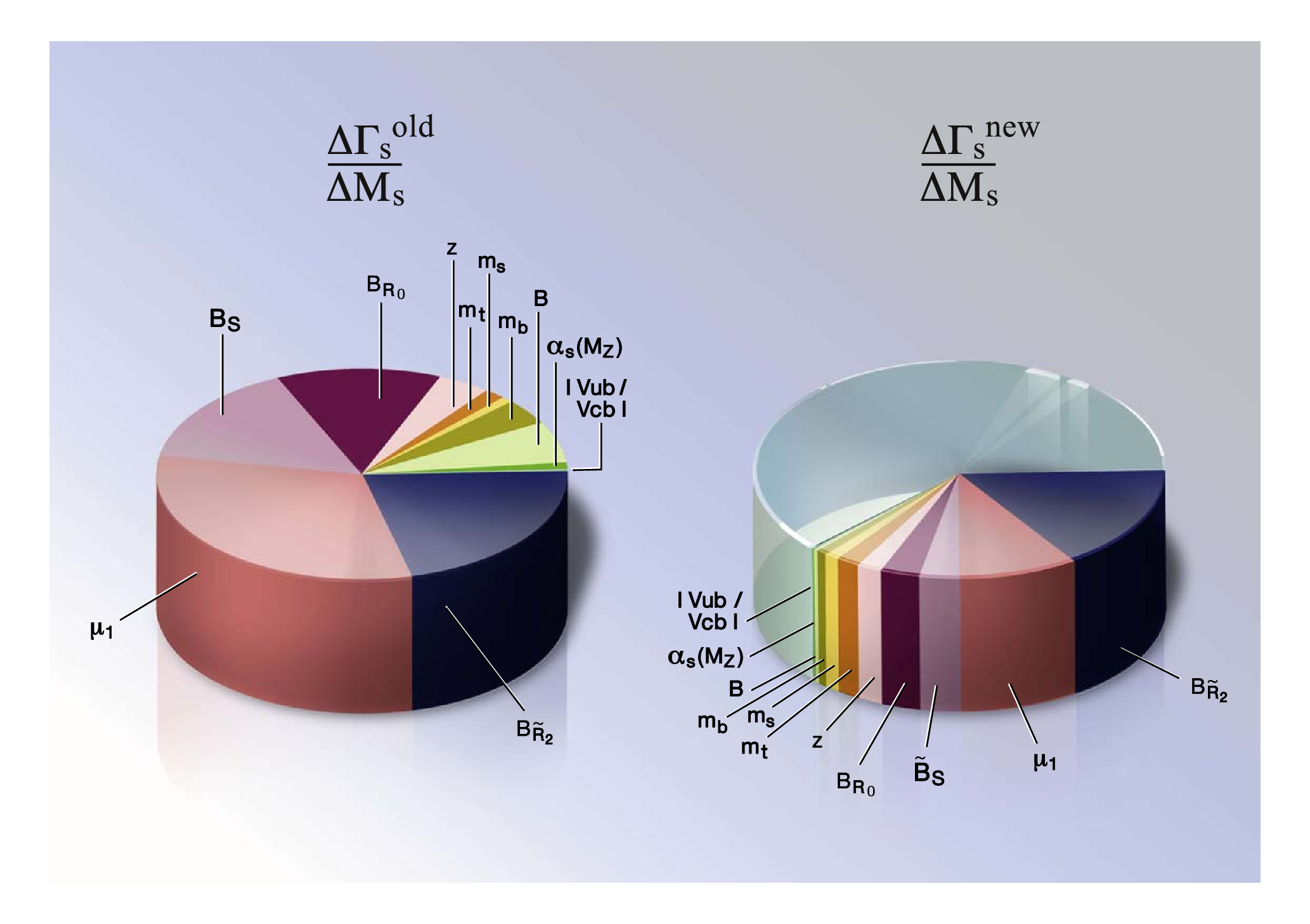,height=4 in}
\caption{Error budget for the determination of $\Delta \Gamma_s / \Delta M_s$ 
in the old and the new basis.}
\label{fig:Kuchen2}
\end{center}
\end{figure}

This changes if one looks at the composition of the theoretical 
error of $\Delta \Gamma / \Delta M$ in Fig. (\ref{fig:Kuchen2}).
Since now $f_{B_s}$ cancels the dominant error comes 
from the uncertainty due to the power 
suppressed operator $\tilde{R}_2$ and the remaining $\mu$-dependence.

One clearly sees that the change of the basis resulted in a considerable
reduction of the theoretical error , 
almost a factor 3 in the case of $\Delta \Gamma_s / \Delta M_s$!
\\
To improve our theoretical knowledge of the mixing quantities further 
one needs more precise values of the non-perturbative parameters, like the 
decay constants or the power suppressed operators.
If accurate non-perturbative parameters are available one might think about 
NNLO calculations ($\alpha_s^2$ or $ \alpha_s/m_b$-corrections) to
reduce the remaining $\mu$-dependence.

\subsection{New Physics}

New physics (see e.g. \cite{newphysics})
is expected to have almost no impact on $\Gamma_{12}$ but it 
can change $M_{12}$ considerably. Therefore one can write
\begin{eqnarray}
\Gamma_{12,s} =  \Gamma_{12,s}^{\rm SM}\, ,
&&
M_{12,s}  =  M_{12,s}^{\rm SM} \cdot { \Delta_s} \, ;
\, \, \, \, \, \, \,  
{ \Delta_s} = { |\Delta_s|} e^{i  \phi^\Delta_s} 
\end{eqnarray}
With this parameterisation the physical mixing parameters can be written as
\begin{eqnarray}
 \Delta M_s  & = & 2 | M_{12,s}^{\rm SM} | \cdot { |\Delta_s |} 
\label{bounddm}
\\
\Delta \Gamma_s  & = & 2 |\Gamma_{12,s}|
\cdot \cos \left( \phi_s^{\rm SM} + { \phi^\Delta_s} \right)
\label{bounddg}
\\
\frac{\Delta \Gamma_s}{\Delta M_s} 
&= &
 \frac{|\Gamma_{12,s}|}{|M_{12,s}^{\rm SM}|} 
\cdot \frac{\cos \left( \phi_s^{\rm SM} + { \phi^\Delta_s} \right)}
{ |\Delta_s|}
\label{bounddgdm}
\\
a_{fs}^s 
&= &
 \frac{|\Gamma_{12,s}|}{|M_{12,s}^{\rm SM}|} 
\cdot \frac{\sin \left( \phi_s^{\rm SM} + { \phi^\Delta_s} \right)}
{ |\Delta_s|}
\label{boundafs}
\end{eqnarray}

Now we combine the current experimental knowledge
about the mixing parameters to find out whether $B_s$-mixing
is described by the standard model alone, or whether we already 
get some signals of new physics contributions.
\\
The mass difference $\Delta M_s$  is now known very precisely 
\cite{menzemer,deltamsexp}
\begin{eqnarray}
\Delta M_s & =&  17.77\pm{0.10} {}_{\mbox{\scriptsize (syst)}} 
       \pm 0.07\,{}_{\mbox{\scriptsize (stat)}} \, \mbox{ps}^{-1} \, 
   \qquad \mbox{CDF} . 
\end{eqnarray} 
For the remaining mixing parameters in the $B_s$-system
only experimental bounds are available.
The width difference $\Delta \Gamma_s /\Gamma_s $ was 
investigated at ALEPH, BELLE, CDF, D0
by analyzing the decays $B_s \to D_s^{(*)+} + D_s^{(*)-} $ \cite{dgexp1},
$B_s \to J / \Psi + \phi $ \cite{dgexp2} and 
$B_s \to K^+ + K^-$ \cite{dgexp3}
and the flavor specific lifetime of the $B_s$ meson \cite{dgexp4}.
A recent combination of all these results yields \cite{dgexpall}
\begin{eqnarray}
\Delta \Gamma_s & = & 0.097 \pm 0.042 \, \mbox{ps}^{-1}\, ,
\\
\frac{\Delta \Gamma_s}{\Delta M_s} & = & 
\left( 56 \pm 24 \right) \times 10^{-4} \, .
\label{eqdgexp}
\end{eqnarray}
Except the angular analysis $B_s \to J / \Psi + \phi $ all other 
determinations are affected by some drawbacks, described in \cite{LN}.
Moreover the D0 collaboration has updated their results 
\cite{dgexp2} for the decay $B_s \to J / \Psi + \phi $ in 
\cite{dgexpnew,ay,menzemer} using 1fb$^-1$ of data.
Setting the value of the mixing phase $\phi_s$ to zero they obtain 
\cite{dgexpnew,ay,menzemer}
\begin{eqnarray}
\Delta \Gamma_s & = & 0.12 \pm 0.08 ^{+0.03}_{-0.04} \, \mbox{ps}^{-1}
\, ,
\label{eqdmexp3}
\end{eqnarray}
allowing for a non-zero value of the mixing phase $\phi_s$ they get
\begin{eqnarray}
\Delta \Gamma_s & = & 0.17 \pm 0.09 \pm 0.03 \, \mbox{ps}^{-1} \, ,
\label{eqdgexp4a}
\\
\phi_s & = & -0.79  \pm 0.56 \pm 0.01 \, .
\label{eqdgexp4b}
\end{eqnarray}
In the following we will for $\Delta \Gamma_s$ and $\phi_s$  
only use the numbers 
from Eq. (\ref{eqdgexp4a}) and Eq. (\ref{eqdgexp4b}).

\begin{figure}%{tb}
\includegraphics[width=0.9\textwidth,angle=0]{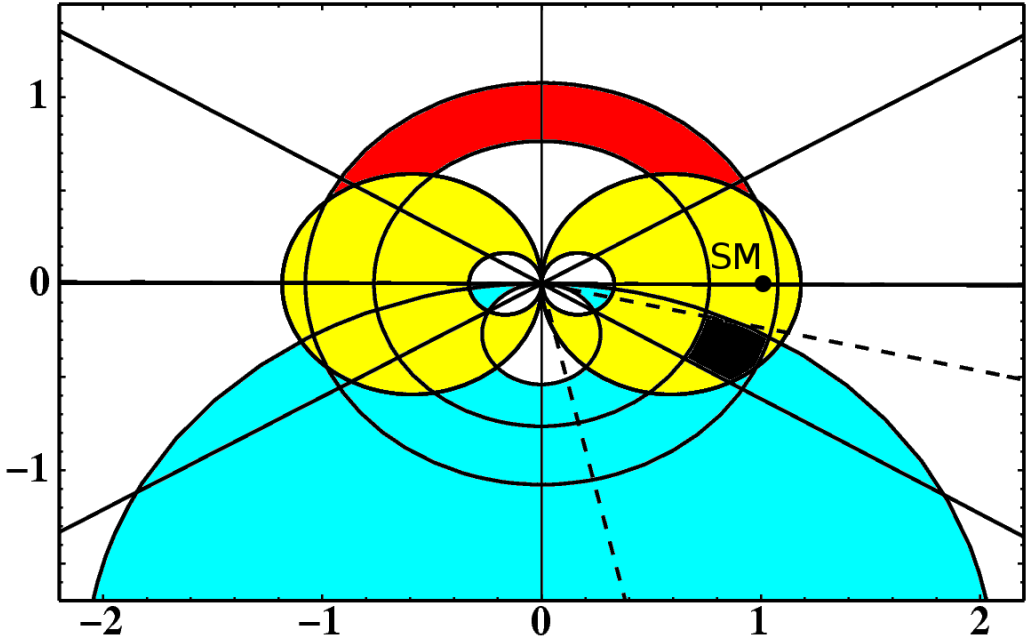}
\caption{Current experimental bounds in the complex $\Delta_s$-plane.
  The bound from $\Delta M_s$ is given by the red (dark-grey) ring around
  the origin. The bound from $\Delta \Gamma_s / \Delta M_s$ is given
  by the yellow (light-grey) region and the bound from $a_{fs}^s$ is given
  by the light-blue (grey) region. The angle $\phi_s^\Delta$ can be extracted
  from $\Delta \Gamma_s$ (solid lines) with a four fold ambiguity - one bound
  coincides with the x-axis! - or from the angular analysis in 
  $B_s \to J / \Psi \phi$ (dashed line). If the standard model is valid 
  all bounds should coincide in the point (1,0). The current experimental 
  situation shows a small deviation, which might become significant, if the 
  experimental uncertainties in $\Delta \Gamma_s$, $a_{sl}^s$ and $\phi_s$ 
  will go down in near future.}\label{boundbandreal}
\end{figure}

The semileptonic CP asymmetry in the $B_s$ system has been determined 
directly in
\cite{aslsexp} and found to be
\begin{equation}
a_{sl}^{s, \rm direct} = \left(24.5 \pm 19.3 \pm 3.5 \right) \cdot 10^{-3} \, .
\label{eqaslexp1}
\end{equation}
Moreover the semileptonic CP asymmetry can be extracted from the
same sign dimuon asymmetry that was measured in \cite{dimuonexp} to be
\begin{equation}
a_{sl} = \left( -2.8 \pm 1.3 \pm 0.9 \right) \cdot 10^{-3} \, .
\label{eqdimuexp}
\end{equation}
Updating the numbers in \cite{combined, nir2006} one sees that
\begin{equation}
a_{sl} = \left(0.582 \pm 0.030 \right) \, a_{sl}^d + 
         \left(0.418 \pm 0.047 \right) \, a_{sl}^s
\label{asllk}
\end{equation}
In \cite{combined, nir2006} the experimental bound for $a_{sl}^d$ was 
used to extract from Eq.(\ref{eqdimuexp}) and Eq.(\ref{asllk}) 
a bound on $a_{sl}^s$. Due to the huge experimental uncertainties in
$a_{sl}^d$ this strategy resulted in a large error on $a_{sl}^s$. Since 
in the $B_d$-system there is not much room left for new physics contributions,
we think it is justified to use the theoretical number of $a_{sl}^d$.
Using $a_{sl}^d = - \left( 0.48 \pm 0.12 \right) \cdot 10^{-3}$ we get from
Eq.(\ref{eqdimuexp}), Eq.(\ref{asllk})  and Eq.(\ref{eqdimuexp}) already 
a nice bound
\begin{equation}
a_{sl}^{s, \rm dimuon} = \left(- 6.0 \pm 3.2 \pm 2.2 \right) \cdot 10^{-3} \, .
\label{eqasldimuon}
\end{equation}
Combining this number with the direct determination \cite{aslsexp}
we get our final experimental number for the semileptonic CP asymmetries
\begin{equation}
a_{sl}^{s} = \left(- 5.2 \pm 3.2 \pm 2.2 \right) \cdot 10^{-3} \, .
\label{eqasldfinal}
\end{equation}
Now we combine these experimental numbers with the theoretical errors
to extract bounds in the imaginary $\Delta_s$-plane by the use of
Eqs. (\ref{bounddm}),  (\ref{bounddg}), (\ref{bounddgdm}) and (\ref{boundafs}),
see Fig. (\ref{boundbandreal}).

The comparison of experiment and standard model expectation for
$\Delta M_s$, $\Delta \Gamma_s$, $\phi_s$, $ \Delta \Gamma_s / \Delta M_s$ 
and $a_{sl}^s$  presented in figure \ref{boundbandreal} already shows some 
hints for deviations from the standard model. 
%

%%%%%%%%%%%%%%%%%%%%%%%%%%%%%%%%%%%%%%%%%%%%%%%%%%%%%%%%%%%%%%%%%%%%%%%%%%%

\section{Conclusion and outlook} 

Theoretical predictions of the lifetimes of heavy mesons are in excellent 
agreement with the experimental numbers.
We do not see any signal of possible duality violations. To become more 
quantitative in the prediction of $\tau_{B^+}/\tau_{B_d}$ 
the non-perturbative estimates of the bag parameters $B_1$, $B_2$ 
and $\epsilon_1$, $\epsilon_2$ have to be improved.
For $\tau_{B_s} / \tau_{B_d}$ more precise experimental numbers are needed, 
while in the case of $\tau_{B_c}$ theoretical progress is mandatory.
\\
The theoretical uncertainty in the mixing parameter $\Delta M$ 
is completely dominated by the decay  constant.
We have presented a method (see \cite{LN} for more details) to reduce the 
theoretical error in $\Delta \Gamma$, $\Delta \Gamma / \Delta M$  
and $a_{fs}$ considerably.
This relatively clean standard model predictions can be used to look for 
new physics effects in $B_s$-mixing. From the currently available experimental 
bounds on $\Delta \Gamma_s$ and $a_{fs}$ one already gets some hints for
deviations from the standard model.
This situation will improve dramatically as soon as more data are available.

For a further reduction of the theoretical uncertainty
in $\Delta \Gamma $ a much higher accuracy than currently available 
on the decay constants is necessary. 
If this problem is solved or if one looks at
quantities like $\Delta \Gamma / \Delta M$ and $a_{fs}$ where 
the dependence on $f_B$ cancels, then the dominant uncertainty 
comes from the unknown matrix elements of the power suppressed operators.
Here any non-perturbative estimate would be very desirable.
If accurate non-perturbative parameters are available one might think about 
NNLO calculations ($\alpha_s^2$ or $ \alpha_s/m_b$-corrections) to
reduce the remaining $\mu$-dependence.

\bigskip
I would like to thank the organizers of HQL2006 for the invitation and
Uli Nierste for the pleasant collaboration.


\begin{thebibliography}{99}

%%
%%  bibliographic items can be constructed using the LaTeX format in SPIRES:
%%    see    http://www.slac.stanford.edu/spires/hep/latex.html
%%  SPIRES will also supply the CITATION line information; please include it.
%%
%%%%%%%%%%%%%%%%%%%%%%%%%%%%%%%%%%%%%%%%%%%%%%%%%%%%%%%%%%%%%%%%%%%%%%%%%
%%%%%%%%%%%%%%%%% Intro %%%%%%%%%%%%%%%%%%%%%%%%%%%%%%%%%%%%%%%%%%%%%%%%%
%%%%%%%%%%%%%%%%%%%%%%%%%%%%%%%%%%%%%%%%%%%%%%%%%%%%%%%%%%%%%%%%%%%%%%%%%
\bibitem{inclusive}
  A.~Lenz, U.~Nierste and G.~Ostermaier,
  %``Penguin diagrams, charmless B decays and the *missing charm puzzle*,''
  Phys.\ Rev.\ D {\bf 56} (1997) 7228
  [arXiv:hep-ph/9706501];
  %%CITATION = HEP-PH 9706501;%%
  A.~Lenz, U.~Nierste and G.~Ostermaier,
  % ``Determination of the CKM angle gamma and |V(ub)/V(cb)| from inclusive
  %direct CP asymmetries and branching ratios in charmless B decays,''
  Phys.\ Rev.\ D {\bf 59} (1999) 034008
  [arXiv:hep-ph/9802202];
  %%CITATION = HEP-PH 9802202;%%
  A.~Lenz,
  %``Some comments on the missing charm puzzle,''
  arXiv:hep-ph/0011258;
  %%CITATION = HEP-PH 0011258;%%
  O. Buchm{\"u}ller, these proceedings;
  T. Mannel, these proceedings.

\bibitem{CERN}
  M.~Battaglia {\it et al.},
  %``The CKM matrix and the unitarity triangle,''
  arXiv:hep-ph/0304132.
  %%CITATION = HEP-PH 0304132;%%

\bibitem{HQE1}
M.~A.~Shifman and M.~B.~Voloshin, in: \emph{Heavy Quarks}\ ed.\
V.~A.~Khoze and M.~A.~Shifman,
%``Heavy Quarks,''
Sov.\ Phys.\ Usp.\  {\bf 26} (1983) 387;
%%CITATION = SOPUA,26,387;%%
M.~A.~Shifman and M.~B.~Voloshin,
%``Preasymptotic Effects In Inclusive Weak Decays Of Charmed Particles,''
Sov.\ J.\ Nucl.\ Phys.\  {\bf 41} (1985) 120
[Yad.\ Fiz.\  {\bf 41} (1985) 187];
%%CITATION = SJNCA,41,120;%%
M.~A.~Shifman and M.~B.~Voloshin,
%``Hierarchy Of Lifetimes Of Charmed And Beautiful Hadrons,''
Sov.\ Phys.\ JETP {\bf 64} (1986) 698
[Zh.\ Eksp.\ Teor.\ Fiz.\  {\bf 91} (1986) 1180];
%%CITATION = SPHJA,64,698;%%

\bibitem{HQE2}
J.~Chay, H.~Georgi and B.~Grinstein,
%``Lepton Energy Distributions In Heavy Meson Decays From QCD,''
Phys.\ Lett.\ B {\bf 247} (1990) 399;
%
I.~I.~Bigi, N.~G.~Uraltsev and A.~I.~Vainshtein,
%``Nonperturbative corrections to inclusive 
%beauty and charm decays: QCD versus phenomenological models,''
Phys.\ Lett.\ B {\bf 293} (1992) 430
[Erratum-ibid.\ B {\bf 297} (1992) 477];
  I.~I.~Y.~Bigi, M.~A.~Shifman, N.~G.~Uraltsev and A.~I.~Vainshtein,
  %``QCD predictions for lepton spectra in inclusive heavy flavor decays,''
  Phys.\ Rev.\ Lett.\  {\bf 71} (1993) 496
  [arXiv:hep-ph/9304225];
  %%CITATION = HEP-PH 9304225;%%
  B.~Blok, L.~Koyrakh, M.~A.~Shifman and A.~I.~Vainshtein,
   ``Differential distributions in semileptonic decays of the heavy flavors in
  %QCD,''
  Phys.\ Rev.\ D {\bf 49} (1994) 3356
  [Erratum-ibid.\ D {\bf 50} (1994) 3572]
  [arXiv:hep-ph/9307247];
  %%CITATION = HEP-PH 9307247;%%
  A.~V.~Manohar and M.~B.~Wise,
  %``Inclusive semileptonic B and polarized Lambda(b) decays from QCD,''
  Phys.\ Rev.\ D {\bf 49} (1994) 1310
  [arXiv:hep-ph/9308246].
  %%CITATION = HEP-PH 9308246;%%

\bibitem{spectator}
  B.~Guberina, S.~Nussinov, R.~D.~Peccei and R.~Ruckl,
  %``D Meson Lifetimes And Decays,''
  Phys.\ Lett.\ B {\bf 89} (1979) 111;
  %%CITATION = PHLTA,B89,111;%%
  N.~Bilic, B.~Guberina and J.~Trampetic,
  %``Pauli Interference Effect In D+ Lifetime,''
  Nucl.\ Phys.\ B {\bf 248} (1984) 261;
  %%CITATION = NUPHA,B248,261;%%
  B.~Guberina, R.~Ruckl and J.~Trampetic,
  %``CHARMED BARYON LIFETIME DIFFERENCES,''
  Z.\ Phys.\ C {\bf 33} (1986) 297.
  %%CITATION = ZEPYA,C33,297;%%


\bibitem{Vittorio}
V.~Lubicz, these proceedings.

\bibitem{LN}
  A.~Lenz and U.~Nierste,
  %``Theoretical update of Bs-Bs-bar mixing,''
  arXiv:hep-ph/0612167.
  %%CITATION = HEP-PH 0612167;%%

\bibitem{Onogi}
  T.~Onogi,
  %``Heavy flavor physics from lattice QCD,''
  arXiv:hep-lat/0610115.
  %%CITATION = HEP-LAT 0610115;%%

\bibitem{updates}
  A.~Lenz,
  % ``Next-to-leading order QCD corrections to the lifetime difference of  B/s
  %mesons,''
  arXiv:hep-ph/9906317;
  %%CITATION = HEP-PH 9906317;%%
  M.~Beneke and A.~Lenz,
  %``Lifetime difference of B/s mesons: Theory status,''
  J.\ Phys.\ G {\bf 27} (2001) 1219
  [arXiv:hep-ph/0012222];
  %%CITATION = HEP-PH 0012222;%%
  A.~Lenz and S.~Willocq,
  %``Mixing and lifetimes summary,''
  J.\ Phys.\ G {\bf 27} (2001) 1207;
  %%CITATION = JPHGB,G27,1207;%%
  A.~Lenz,
  % ``Theoretical status of the lifetime predictions:  
  %(Delta(Gamma)/Gamma)(B/s),
  %tau(B+)/tau(B/d) and tau(Lambda/b)/tau(B/d),''
  arXiv:hep-ph/0107033;
  %%CITATION = HEP-PH 0107033;%%
  A.~Lenz,
  %``Decay rate difference in the neutral B system: Delta(Gamma(B/s)) and
  %Delta(Gamma(B/d)),''
  arXiv:hep-ph/0412007.
  %%CITATION = HEP-PH 0412007;%%

%%%%%%%%%%%%%%%%%%%%%%%%%%%%%%%%%%%%%%%%%%%%%%%%%%%%%%%%%%%%%%%%%%%%%%%%%
%%%%%%%%%%%%%%%  Lifetimes  %%%%%%%%%%%%%%%%%%%%%%%%%%%%%%%%%%%%%%%%%%%%%
%%%%%%%%%%%%%%%%%%%%%%%%%%%%%%%%%%%%%%%%%%%%%%%%%%%%%%%%%%%%%%%%%%%%%%%%%
\bibitem{tauLO}
  I.~I.~Y.~Bigi, B.~Blok, M.~A.~Shifman, N.~Uraltsev and A.~I.~Vainshtein,
  %``Nonleptonic decays of beauty hadrons: From phenomenology to theory,''
  arXiv:hep-ph/9401298;
  %%CITATION = HEP-PH 9401298;%%
  I.~I.~Y.~Bigi,
  %``The QCD Perspective On Lifetimes Of Heavy Flavor Hadrons,''
  arXiv:hep-ph/9508408;
  %%CITATION = HEP-PH 9508408;%%
  M.~Neubert and C.~T.~Sachrajda,
  %``Spectator effects in inclusive decays of beauty hadrons,''
  Nucl.\ Phys.\ B {\bf 483} (1997) 339
  [arXiv:hep-ph/9603202].
  %%CITATION = HEP-PH 9603202;%%


\bibitem{BBGLN02}
  M.~Beneke, G.~Buchalla, C.~Greub, A.~Lenz and U.~Nierste,
  %``The B+ - B/d0 lifetime difference beyond leading logarithms,''
  Nucl.\ Phys.\ B {\bf 639} (2002) 389
  [arXiv:hep-ph/0202106].
  %%CITATION = HEP-PH 0202106;%%

\bibitem{rome02}
  E.~Franco, V.~Lubicz, F.~Mescia and C.~Tarantino,
  %``Lifetime ratios of beauty hadrons at the next-to-leading order in QCD,''
  Nucl.\ Phys.\ B {\bf 633} (2002) 212
  [arXiv:hep-ph/0203089].
  %%CITATION = HEP-PH 0203089;%%


\bibitem{lifetime1m}
A.~Lenz and U.~Nierste, unpublished, e.g. talk at Academia Sinica 2003;
F.~Gabbiani, A.~I.~Onishchenko and A.~A.~Petrov,
  %``Spectator effects and lifetimes of heavy hadrons,''
  Phys.\ Rev.\ D {\bf 70} (2004) 094031
  [arXiv:hep-ph/0407004].
  %%CITATION = HEP-PH 0407004;%%


\bibitem{lifetimelattice}
  M.~Di Pierro and C.~T.~Sachrajda  [UKQCD Collaboration],
  %``A lattice study of spectator effects in inclusive decays of B mesons,''
  Nucl.\ Phys.\ B {\bf 534} (1998) 373
  [arXiv:hep-lat/9805028];
  %%CITATION = HEP-LAT 9805028;%%
  D.~Becirevic,
  %``Theoretical progress in describing the B meson lifetimes,''
  arXiv:hep-ph/0110124.
  %%CITATION = HEP-PH 0110124;%%

\bibitem{fB}
  A.~Gray {\it et al.}  [HPQCD Collaboration],
  %``The B meson decay constant from unquenched lattice QCD,''
  %fBd = 216pm22, fBs/fbd = 1.20pm 0.03, fBs = 259pm32
  Phys.\ Rev.\ Lett.\  {\bf 95} (2005) 212001
  [arXiv:hep-lat/0507015].
  %%CITATION = HEP-LAT 0507015;%%


\bibitem{PDG}
  W.~M.~Yao {\it et al.}  [Particle Data Group],
  %``Review of particle physics,''
  J.\ Phys.\ G {\bf 33} (2006) 1.
  %%CITATION = JPHGB,G33,1;%%

\bibitem{ay}
A. Cano, these proceedings.

\bibitem{tauBs} 
  Y.~Y.~Keum and U.~Nierste,
  %``Probing penguin coefficients with the lifetime ratio tau(B/s)/tau(B/d),''
  Phys.\ Rev.\ D {\bf 57} (1998) 4282
  [arXiv:hep-ph/9710512].
  %%CITATION = HEP-PH 9710512;%%

\bibitem{BBD}
  M.~Beneke, G.~Buchalla and I.~Dunietz,
  %``Width Difference in the $B_s-\bar{B_s}$ System,''
  Phys.\ Rev.\ D {\bf 54} (1996) 4419
  [arXiv:hep-ph/9605259].
  %%CITATION = HEP-PH 9605259;%%

\bibitem{HFAG}
  http://www.slac.stanford.edu/xorg/hfag/index.html

\bibitem{taubctheory}
  C.~H.~Chang, S.~L.~Chen, T.~F.~Feng and X.~Q.~Li,
  %``The lifetime of B/c meson and some relevant problems,''
  % 0.36 ps
  Phys.\ Rev.\ D {\bf 64} (2001) 014003
  [arXiv:hep-ph/0007162];
  %%CITATION = HEP-PH 0007162;%%
  V.~V.~Kiselev, A.~E.~Kovalsky and A.~K.~Likhoded,
  %``B/c decays and lifetime in QCD sum rules,''
  % 0.48 pm 0.05
  Nucl.\ Phys.\ B {\bf 585} (2000) 353
  [arXiv:hep-ph/0002127].
  %%CITATION = HEP-PH 0002127;%%
  A.~Y.~Anisimov, I.~M.~Narodetsky, C.~Semay and B.~Silvestre-Brac,
  %``The B/c meson lifetime in the light-front constituent quark model,''
  % 0.59 pm 0.06
  Phys.\ Lett.\ B {\bf 452} (1999) 129
  [arXiv:hep-ph/9812514].
  %%CITATION = HEP-PH 9812514;%%
  M.~Beneke and G.~Buchalla,
  %``The $B_c$ Meson Lifetime,''
  % 0.4 ... 0.7
  Phys.\ Rev.\ D {\bf 53} (1996) 4991
  [arXiv:hep-ph/9601249].
  %%CITATION = HEP-PH 9601249;%%

\bibitem{taubcexp}
  A.~Abulencia {\it et al.}  [CDF Collaboration],
  % ``Measurement of the $B_c^+$ meson lifetime using $B_c^+ \to J/\psi e^+
  %\nu_e$,''
  % 0.463 + 0.073 - 0.065 pm 0.036: 360 fb^-1
  Phys.\ Rev.\ Lett.\  {\bf 97} (2006) 012002
  [arXiv:hep-ex/0603027].
  %%CITATION = HEP-EX 0603027;%%
  D0 Collaboration, D0 conference note 4539
  % 0.448 + 0.123 - 0.096 pm 0.121

%%%%%%%%%%%%%%%%%%%%%%%%%%%%%%%%%%%%%%%%%%%%%%%%%%%%%%%%%%%%%%%%%%%%%%%%%
%%%%%%%%%%%%%%% Mixing   %% %%%%%%%%%%%%%%%%%%%%%%%%%%%%%%%%%%%%%%%%%%%%%
%%%%%%%%%%%%%%%%%%%%%%%%%%%%%%%%%%%%%%%%%%%%%%%%%%%%%%%%%%%%%%%%%%%%%%%%%
\bibitem{IL}
  T.~Inami and C.~S.~Lim,
  % ``Effects Of Superheavy Quarks And Leptons 
  %In Low-Energy Weak Processes K(L)
  %$\to$ Mu Anti-Mu, K+ $\to$ Pi+ Neutrino 
  %Anti-Neutrino And K0 <---> Anti-K0,''
  Prog.\ Theor.\ Phys.\  {\bf 65} (1981) 297
  [Erratum-ibid.\  {\bf 65} (1981) 1772].
  %%CITATION = PTPKA,65,297;%%

\bibitem{BJW}
  A.~J.~Buras, M.~Jamin and P.~H.~Weisz,
  % ``LEADING AND NEXT-TO-LEADING QCD CORRECTIONS TO epsilon PARAMETER AND B0 -
  %anti-B0 MIXING IN THE PRESENCE OF A HEAVY TOP QUARK,''
  Nucl.\ Phys.\ B {\bf 347} (1990) 491.
  %%CITATION = NUPHA,B347,491;%%

\bibitem{JLQCD}
 S.~Aoki {\it et al.}  [JLQCD Collaboration],
  %``B0 anti-B0 mixing in unquenched lattice QCD,''
  Phys.\ Rev.\ Lett.\  {\bf 91} (2003) 212001
  [arXiv:hep-ph/0307039].
  %%CITATION = HEP-PH 0307039;%%


\bibitem{menzemer}
S. Menzemer, these proceedings.

\bibitem{deltamsexp}
  A.~Abulencia {\it et al.}  [CDF Collaboration],
  %``Observation of B/s0 anti-B/s0 oscillations,''
  arXiv:hep-ex/0609040;
  %%CITATION = HEP-EX 0609040;%%
A.~Abulencia  [CDF - Run II Collaboration],
%   ``Measurement of the B/s0 anti-B/s0 oscillation frequency,''
 Phys.\ Rev.\ Lett.\  {\bf 97} (2006) 062003
  [arXiv:hep-ex/0606027];
  %%CITATION = HEP-EX 0606027;%%
 V.~M.~Abazov {\it et al.}  [D0 Collaboration],
  %``First direct two-sided bound on the B/s0 oscillation frequency,''
  Phys.\ Rev.\ Lett.\  {\bf 97} (2006) 021802
  [arXiv:hep-ex/0603029].
  %%CITATION = HEP-EX 0603029;%%

\bibitem{dgLO}
 J.S. Hagelin, Nucl. Phys. {\bf B193}, 123 (1981);
 E. Franco, M. Lusignoli and A. Pugliese, 
 Nucl. Phys. {\bf B194}, 403 (1982);
 L.L. Chau, Phys. Rep. {\bf 95}, 1 (1983);
 A.J. Buras, W. S\l ominski and H. Steger, 
 Nucl. Phys. {\bf B245}, 369 (1984);
 M.B. Voloshin, N.G. Uraltsev, V.A. Khoze and M.A. Shifman, 
 Sov. J. Nucl. Phys. {\bf 46}, 112 (1987);
 A. Datta, E.A. Paschos and U. T\"urke, 
 Phys. Lett. {\bf B196}, 382 (1987);
 A. Datta, E.A. Paschos and Y.L. Wu,
 Nucl. Phys. {\bf B311}, 35 (1988).

\bibitem{BBGLN98}
  M.~Beneke, G.~Buchalla, C.~Greub, A.~Lenz and U.~Nierste,
  % ``Next-to-leading order {QCD} corrections to the lifetime 
  % difference of  B/s mesons,''
  Phys.\ Lett.\ B {\bf 459} (1999) 631
  [arXiv:hep-ph/9808385].
  %%CITATION = HEP-PH 9808385;%%

\bibitem{BBLN03}
  M.~Beneke, G.~Buchalla, A.~Lenz and U.~Nierste,
  %``CP asymmetry in flavour-specific B decays beyond leading logarithms,''
  Phys.\ Lett.\ B {\bf 576} (2003) 173
  [arXiv:hep-ph/0307344].
  %%CITATION = HEP-PH 0307344;%%

\bibitem{rome03}
  M.~Ciuchini, E.~Franco, V.~Lubicz, F.~Mescia and C.~Tarantino,
  % ``Lifetime differences and CP violation parameters of neutral B mesons at
  %the next-to-leading order in QCD,''
  JHEP {\bf 0308} (2003) 031
  [arXiv:hep-ph/0308029].
  %%CITATION = HEP-PH 0308029;%%


\bibitem{1overm}
  A.~S.~Dighe, T.~Hurth, C.~S.~Kim and T.~Yoshikawa,
  % ``Measurement of the lifetime difference of B/d mesons: Possible and
  %worthwhile?,''
  Nucl.\ Phys.\ B {\bf 624} (2002) 377
  [arXiv:hep-ph/0109088].
  %%CITATION = HEP-PH 0109088;%%

\bibitem{LN2}
A.~Lenz and U.~Nierste, to appear.


\bibitem{MS}
  W.~A.~Bardeen, A.~J.~Buras, D.~W.~Duke and T.~Muta,
  %      \nonumber ``Deep Inelastic Scattering Beyond The Leading Order In Asymptotically Free
  %Gauge Theories,''
  Phys.\ Rev.\ D {\bf 18} (1978) 3998.
  %%CITATION = PHRVA,D18,3998;%%

\bibitem{Becirevic}
D.~Becirevic, V.~Gimenez, G.~Martinelli, M.~Papinutto and J.~Reyes,
%``B-parameters of the complete set of matrix elements of Delta(B) = 2
%operators from the lattice,''
JHEP {\bf 0204} (2002) 025
[arXiv:hep-lat/0110091].

\bibitem{newphysics}
A. Weiler, these proceedings; 
T. Iijima, these proceedings;
C. Tarantino, these proceedings;
R. Fleischer, these proceedings.

\bibitem{dgexp1}
R.~Barate {\it et al.}  [ALEPH Collaboration],
%   ``A study of the decay width difference in the B/s0 anti-B/s0 system  using
%   Phi Phi correlations,''
%
Phys.\ Lett.\ B {\bf 486} (2000) 286;
%%CITATION = PHLTA,B486,286;%%
CDF collaboration, conference note 7925, http://www-cdf.fnal.gov;
D0 collaboration, conference note 5068, http://www-do.fnal.gov/;
K.~Ikado  [Belle Collaboration],
%``Hot topics from Belle experiment,''
 eConf {\bf C060409} (2006) 003
[arXiv:hep-ex/0605068];
%%CITATION = HEP-EX 0605068;%%
A.~Drutskoy,
%``Results from the Upsilon(5S) engineering run (BELLE),''
arXiv:hep-ex/0605110;
%%CITATION = HEP-EX 0605110;%
M. Paulini, these proceedings.

\bibitem{dgexp2}
D.~Acosta {\it et al.}  [CDF Collaboration],
%   ``Analysis of decay-time dependence of angular distributions in B/s0 $\to$
%   J/psi Phi and B/d0 $\to$ J/psi K*0 decays and measurement of the lifetime
%   difference between B/s mass eigenstates,''
%
Phys.\ Rev.\ Lett.\  {\bf 94} (2005) 101803
[arXiv:hep-ex/0412057];
D0  collaboration, conference note 5052, http://www-do.fnal.gov/.

\bibitem{dgexp3}
  D.~Tonelli  [CDF Collaboration],
  %``CDF hot topics,''
  eConf {\bf C060409} (2006) 001
  [arXiv:hep-ex/0605038];
  %%CITATION = HEP-EX 0605038;%%


A.~Abulencia  [CDF Collaboration],
%``Observation of Bs->K+K- and Measurements of Branching Fractions of
%Charmless Two-body Decays of B0 and Bs Mesons in p-pbar Collisions at
%sqrt(s)=1.96 TeV,''
arXiv:hep-ex/0607021.


\bibitem{dgexp4}
[Heavy Flavor Averaging Group (HFAG)],
%``Averages of b-hadron properties at the end of 2005,''
%
arXiv:hep-ex/0603003;
%%CITATION = HEP-EX 0603003;%%
V.~M.~Abazov {\it et al.}  [D0 Collaboration],
%``A precise measurement of the B/s0 lifetime,''
%
arXiv:hep-ex/0604046.
%%CITATION = HEP-EX 0604046;%%

\bibitem{dgexpall}
  R.~Van Kooten,
  %``B/s0 decays and B hadron leptonic decays,''
  eConf {\bf C060409} (2006) 031
  [arXiv:hep-ex/0606005].
  %%CITATION = HEP-EX 0606005;%%


\bibitem{dgexpnew}
D0 collaboration, conference note 5144, http://www-do.fnal.gov/.

\bibitem{aslsexp}
D0 collaboration, conference note 5143, http://www-do.fnal.gov/.

\bibitem{dimuonexp}
  V.~Abazov  [D0 Collaboration],
   ``Measurement of the CP-violation parameter of B0 mixing and decay with p
  %$\bar{p} \to \mu \mu X$ data,''
  arXiv:hep-ex/0609014.
  %%CITATION = HEP-EX 0609014;%%

\bibitem{combined}
D0 collaboration, conference note 5189, http://www-do.fnal.gov/.

\bibitem{nir2006}
  Y.~Grossman, Y.~Nir and G.~Raz,
  %``Constraining the phase of B/s - anti-B/s mixing,''
  Phys.\ Rev.\ Lett.\  {\bf 97} (2006) 151801
  [arXiv:hep-ph/0605028].
  %%CITATION = HEP-PH 0605028;%%


\end{thebibliography}
\end{document}